\documentclass[preprint, 12pt, a4paper]{elsarticle}

\usepackage{amssymb}
\usepackage{amsmath}
\usepackage{amsthm}
\usepackage{xfrac}
\usepackage[binary-units, per-mode=fraction, fraction-function=\sfrac]{siunitx}

\RequirePackage{xspace} % to get your spacing straight
\DeclareRobustCommand{\Rayleigh}{\ensuremath{R\hspace{-0.1em}a}\xspace}
\DeclareRobustCommand{\Reynolds}{\ensuremath{R\hspace{-0.1em}e}\xspace}

\DeclareRobustCommand{\Rei}{\ensuremath{R\hspace{-0.1em}e_{\text{i}}}\xspace}
\DeclareRobustCommand{\Nui}{\ensuremath{N\hspace{-0.1em}u_{\text{i}}}\xspace}

\RequirePackage{hyperref}
\RequirePackage{url} % to avoid problems with special characters in email and web address
\RequirePackage{xspace} % to get your spacing straight
\DeclareRobustCommand{\blas}{\href{http://www.netlib.org/blas}{\code{BLAS}}\xspace}
\DeclareRobustCommand{\code}[1]{\textbf{\texttt{\color{black}#1}}\xspace}
\DeclareRobustCommand{\c}{\href{https://en.wikipedia.org/wiki/C_(programming_language)}{\code{C}}\xspace}
\DeclareRobustCommand{\cf}{\href{http://www.channelflow.ch}{\code{channelflow}}}
\DeclareRobustCommand{\cuda}{\href{https://developer.nvidia.com/cuda-toolkit}{\code{CUDA}}\xspace}
\DeclareRobustCommand{\fftw}{\href{http://fftw.org/}{\code{FFTW}}\xspace}
\DeclareRobustCommand{\ford}{\href{http://mjr.pages.mpcdf.de/nscouette/ford-doc/}{\code{Ford}}\xspace}
\DeclareRobustCommand{\fortran}{\href{https://en.wikipedia.org/wiki/Fortran}{\code{Fortran}}\xspace}
\DeclareRobustCommand{\hdf}{\href{https://en.wikipedia.org/wiki/Hierarchical_Data_Format}{\code{HDF5}}\xspace}
\DeclareRobustCommand{\lapack}{\href{http://www.netlib.org/lapack}{\code{LAPACK}}\xspace}
\DeclareRobustCommand{\mpi}{\href{https://en.wikipedia.org/wiki/Message_Passing_Interface}{\code{MPI}}\xspace}
\DeclareRobustCommand{\nsc}{\href{https://github.com/dfeldmann/nsCouette}{\code{nsCouette}}\xspace}
\DeclareRobustCommand{\nsp}{\href{https://github.com/dfeldmann/nsCouette}{\code{nsPipe}}\xspace}
\DeclareRobustCommand{\omp}{\href{https://www.openmp.org/}{\code{OpenMP}}\xspace}
\DeclareRobustCommand{\opf}{\href{www.openpipeflow.org}{\code{openpipeflow}}\xspace}
\DeclareRobustCommand{\paraview}{\href{https://en.wikipedia.org/wiki/ParaView}{\code{ParaView}}\xspace}
\DeclareRobustCommand{\python}{\href{https://www.python.org/}{\code{Python}}\xspace}
\DeclareRobustCommand{\visit}{\href{https://en.wikipedia.org/wiki/VisIt}{\code{VisIt}}\xspace}
\DeclareRobustCommand{\xmf}{\href{http://www.xdmf.org}{\code{xdmf}}\xspace}
\DeclareRobustCommand{\gcc}{\href{https://gcc.gnu.org/}{\code{GCC}}\xspace}
\DeclareRobustCommand{\ompi}{\href{https://www.open-mpi.org/}{\code{OpenMPI}}\xspace}
\DeclareRobustCommand{\oblas}{\href{https://www.openblas.net/}{\code{OpenBLAS}}\xspace}

\RequirePackage{lineno}
\RequirePackage{xcolor}
\RequirePackage{layouts}

\journal{SoftwareX}
\begin{document}

% from the authors guide:
% 4. The manuscript must be submitted in single column.
% The following constraints apply: Word count: max. 3000.
% a. Excluding: title, authors, affiliations, references, metadata tables.
% b. Including: abstract, running text, captions, footnotes.
% c. Max. 6 figures.

\begin{frontmatter}

\title{\nsc~-- A high-performance code for direct numerical simulations of turbulent Taylor--Couette flow}

\author[1]{Jose~Manuel~L\'opez\corref{ca2}}
\author[2]{Daniel~Feldmann\corref{ca1}}
\author[3]{Markus~Rampp}
\author[4]{Alberto~Vela-Mart\'in}
\author[5]{Liang~Shi}
\author[2]{Marc~Avila}
\address[1]{Institute of Science and Technology Austria,\\ Am Campus 1, 3400 Klosterneuburg, Austria}
\address[2]{University of Bremen, Center of Applied Space Technology and Microgravity (ZARM),\\ Am Fallturm 2, 28359 Bremen, Germany}
\address[3]{Max Planck Computing and Data Facility (MPCDF),\\ Gie\ss enbachstra\ss e 2, 85748 Garching, Germany}
\address[4]{School of Aeronautics, Universidad Polit\'ecnica de Madrid,\\ Plaza del Cardenal Cisneros 3, 28040 Madrid, Spain}
\address[5]{Max Planck Institute for Dynamics and Self-Organization (MPIDS),\\ Bunsenstra\ss e 10, 37073 G\"ottingen, Germany}
\cortext[ca1]{Corresponding author at ZARM: daniel.feldmann@zarm.uni-bremen.de}
\cortext[ca2]{Corresponding author at IST Austria: jlopez@ist.ac.at}

% 82 words (max ca. 100)
\begin{abstract}
We present \nsc, a highly scalable software tool to solve the Navier--Stokes
equations for incompressible fluid flow between differentially heated and
independently rotating, concentric cylinders. It is based on a pseudospectral
spatial discretization and dynamic time-stepping. It is implemented in modern
\fortran with a hybrid \mpi-\omp parallelization scheme and thus designed to
compute turbulent flows at high Reynolds and Rayleigh numbers. An additional GPU
implementation (\c-\cuda) for intermediate problem sizes and a version for
pipe flow (\nsp) are also provided.
\par\noindent
Link to open access journal publication: \href{https://doi.org/10.1016/j.softx.2019.100395}{DOI: 10.1016/j.softx.2019.100395}
\end{abstract}


\end{frontmatter}

% \linenumbers

\section{Motivation and significance}
\label{sec:motivationAndSignificance}
% 196 Words

Flows in engineering and nature are often characterized by large Reynolds
(\Reynolds) or Rayleigh (\Rayleigh) numbers. Examples are the flow of gas in
astrophysical disks, atmospheric flows and the cooling of rotating machines. In
most cases, it is impossible to resolve all scales of the turbulent flow in a
direct numerical simulation (DNS). However, DNS provide reliable data to allow
extrapolation to the large \Reynolds limit and to enable the development of
adequate subgrid-scale models. Taylor--Couette (TC) flow -- the flow between two
independently rotating concentric cylinders -- stands out as a testbed for these
purposes~\cite{Bazilevs2010, Grossmann2016}. It allows exploring a variety of
physical mechanisms, including buoyancy, shear, rotation and boundary layers in
the vicinity of curved walls. Our DNS code \nsc integrates the incompressible
Navier--Stokes equations for TC flow forward in time using cylindrical
coordinates and primitive variables. Optionally, the cylinder walls can be
differentially heated, in which case an additional equation for the temperature
is solved. The goal of this paper is to make \nsc publicly available and thus
enable DNS of rotating turbulent shear flows to a wide range of users in the
mathematics, physics and engineering communities.

\section{Software description}
\label{sec:softwareDescription}

\subsection{Functionality}
\label{sec:functionality}
% 409 words

In \nsc, the governing equations are discretized using a pseudospectral
Fourier--Galerkin ansatz for the azimuthal ($\theta$) and the axial ($z$)
direction. High-order finite differences (FD) are used in $r$; the only
inhomogeneous direction. The user can select the distribution of
radial grid points at runtime and the stencil-length of the FD scheme is
specified at compile time with a default of nine points.
\par
Periodic boundary conditions (BC) are assumed in $z$ to avoid the need
for dense grids close to the vertical boundaries. The comparison between DNS
with axially periodic BCs and laboratory experiments with solid
end-plates is extremely satisfactory for a wide range of \Reynolds from
laminar to highly turbulent flows~\cite{Grossmann2016}. Additionaly,
$z$-periodicity often provides a more accurate model of astrophysical and
geophysical flows and prevents misleading physical interpretations due to
undesired end-wall-effects~\cite{Edlund2014, Lopez2017}. Details of the method
and implementation are published in~\cite{Shi2015}.
\par
The temporal integration scheme has been upgraded to a predictor-cor\-rec\-tor
method~\cite{Guseva2015}. This enables a variable time-step size with dynamic
control, which is of advantage if the flow state is suddenly modified (applying
disturbances, changing rotation rates) or naturally undergoes strongly transient
dynamics.
\par
Another significant upgrade is the extension to heat transfer where a
temperature difference between the cylinders is imposed. To this end, a
Boussinesq-like approximation~\cite{Lopez2013} has been implemented to account
for buoyancy effects. In the distributed version of \nsc, a negative temperature
gradient in $r$ is considered, whereas gravity is aligned in $z$. Other
scenarios can be easily investigated by changing only a few lines of source
code.
\par
Additionally, divergence-free initial conditions can be used to easily excite
selected Fourier modes. This enables the user to
systematically investigate different transition scenarios.
\par
A single input file defines all relevant parameters (number of points, modes and
timesteps, rotation rates etc.) at runtime. At every restart, the spatial
resolution can be changed using an automated interpolation and mode
padding functionality.
\par
This and many other convenient features, together with a number of example
\code{Makefiles} for the most common high-performance computing (HPC) platforms,
provide newcomers an easy start into the world of highly-resolved and
massively-parallel DNS. A user guide is included to help non-expert
users get started with \nsc (compilation, setup, select proper resolution,
analyse data,
etc.).

\subsection{Software architecture}
\label{sec:softwareArchitecture}
% 476 words

Over time, \nsc has been ported to all major CPU-based HPC platforms. Amongst
IBM Power, BlueGene and \code{x86\_64} architectures -- including a few
generations of the prevalent Intel Xeon multi-core processors -- it has also
been ported to Xeon Phi (KnightsLanding), AMD EPYC (Naples, Rome) and
ARMv8.1 (Marvell ThunderX2) platforms. Developments for the NEC SX-Aurora vector
architecture and multi-GPU clusters are underway.
\par
Building the executable requires a modern \fortran compiler, a standard \c
compiler and only very few additional libraries: \mpi, \blas/\lapack, \fftw and
optionally \hdf. All of them are commonly available as high-quality, open-source
software (e.g. \gcc, \ompi~\cite{Gabriel2004}, \fftw~\cite{Frigo2005}, \oblas)
and as vendor-optimized tool chains (e.g. Intel Parallel Studio \code{PSXE}).
\par
Our code runs on laptops and -- for large-enough problems -- efficiently scales
up to the largest HPC systems with tens of thousands of processor
cores~\cite{Rampp2014}. The basic architecture of \nsc and our design
choice for a hybrid \mpi-\omp parallelization scheme (HPS) follows
straightforwardly from the Fourier--Galerkin ansatz and from the current
technological trend of multi-core processors with ever increasing core
counts and stagnating per-core performance. The basis of our HPS is a
one-dimensional (1d) \mpi-only slab decomposition into Fourier modes, which can
be treated independently of each other in the computation of the linear terms
occurring in the governing equations. For computing the non-linear terms, global
data transpositions (\code{MPI\_Alltoall}) and task-local transposes are
employed to gather all modes locally on each \mpi task. Therefore, the
number of radial grid points ($N_r$), which is typically much smaller than the
number of Fourier modes, imposes a natural limit to the 1d \mpi-only approach.
We relax this limit by introducing an additional parallelization layer, such
that multiple cores per \mpi task will be allocated to accommodate multiple \omp
threads. First, this allows to compute the linear terms in \omp-parallel loops
over Fourier modes. Second, an \omp-coarse-grain parallelism can be exploited
during the computations of the non-linear terms by overlapping the global
transpositions and the Fourier transformations of state variables. Overall,
when confronted with a 2d \mpi-only domain decomposition (e.g.~\cite{Willis2017}),
which might be superior for certain setups, our HPS is an excellent compromise
between versatility, simplicity of the implementation and achievable peak parallel
scalability.
\par
Additionally, we provide a basic GPU version of \nsc written in \c-\cuda,
which implements the exact same numerical schemes as the \fortran
version, but currently provides less output and functionalities. These will
be added in the future. A \cuda-capable GPU device with compute capability
2.0 (or higher), support for double-precision arithmetic, and NVIDIA's \cuda
toolkit are required. The GPU version runs on single GPU devices in
a massively parallel setup with thousands of GPU threads and highly efficient
memory management. It relies on custom \cuda kernels for linear algebra,
uses the highly-optimised \code{cuFFT} library to perform Fourier
transforms without the need for inter-node communication and, therefore,
provides a clear speedup with respect to the \fortran code for small problems
that fit into the main memory (RAM) of one GPU.

\subsection{Computational performance}
\label{sec:computationalPerformance}
% 753 words

\begin{figure}[htb]
\centering
\includegraphics[width=1.00\columnwidth]{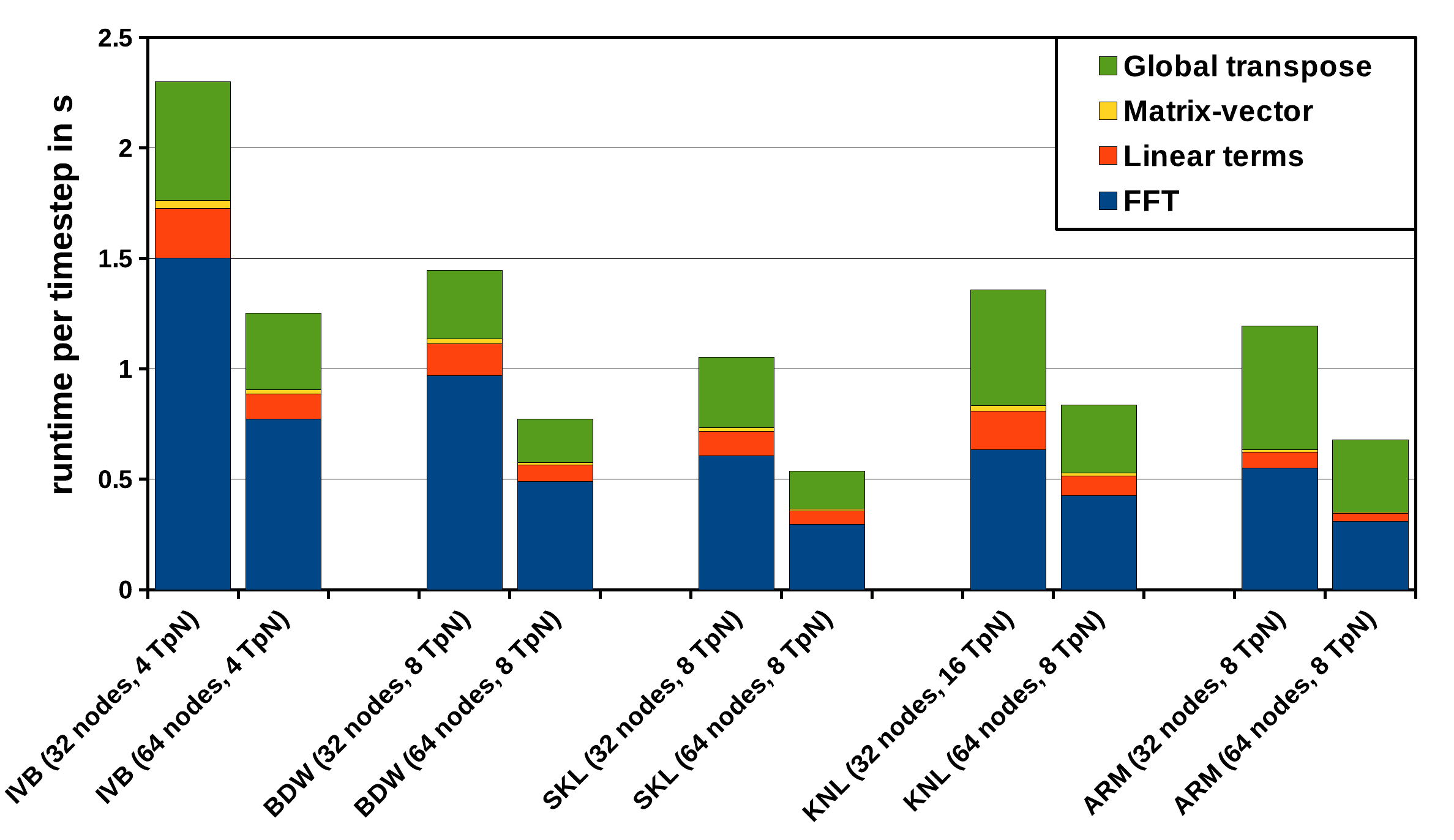}
\caption{Runtime per timestep and breakdown into the main algorithmic components
(different colours) of a typical \nsc run ($N_r=\num{512}$ and
$\num{513}\times\num{1025}$ Fourier modes) computed on \num{32} and \num{64}
dual-socket nodes of various HPC clusters, using a platform-specific number of
\mpi tasks/node (TpN).
IVB: Intel Xeon E5-2680v2 (IvyBridge), \num{20} cores/node.
BDW: Intel Xeon E5-2698v4 (Broadwell), \num{40} cores/node.
SKL: Intel Xeon 6148 (Skylake), \num{40} cores/node.
KNL: Xeon Phi 7230 (KnightsLanding), \num{64} cores/node.
ARM: Marvell ThunderX2 ARM v8.1, \num{64} cores/node.
The IVB and BDW clusters employ a Mellanox InfiniBand FDR network
(\SI{56}{\giga\bit\per\second}), whereas SKL and KNL use Intel OmniPath
(\SI{100}{\giga\bit\per\second}). The ARM cluster is interconnected with Cray
Aries (\SI{80}{\giga\bit\per\second}). \nsc was built using platform-optimized
software tool chains (i.e. compilers and libraries) but no platform-specific
optimization of the source code was performed. Corresponding \code{Makefiles}
are shipped with the code.}
\label{fig:runtime}
\end{figure}

The number of \mpi tasks can be selected at program start-up and changed between
runs with the only restriction that it must divide $N_{r}$. The HPS of \nsc
achieves a scalability well beyond the limitations imposed by 1d \mpi-only
approaches and maps naturally to the prevailing multi-node, multi-core HPC
architectures. Specifically, the flexibility to choose appropriate combinations
for the numbers of \mpi tasks per node and \omp threads per \mpi task has proven
key for achieving good performance across a wide range of architectures from
\num{16} to \num{64} cores per node. On such machines, \nsc enables highly
resolved DNS with $N_{r}=\mathcal{O}(\num{e3})$ using
$\mathcal{O}(\num{e4})$ cores~\cite{Shi2015}. In absolute terms, for an
intermediate problem size with $N_{r}=\num{512}$ and $\num{513}\times\num{1025}$
Fourier modes (i.e. \Reynolds up to $\mathcal{O}(\num{e4})$), the computation of
a single timestep takes less than a second on \num{64} nodes (\num{2560} cores)
of a contemporary HPC cluster (Figure~\ref{fig:runtime}) and requires roughly
\SI{450}{\giga\byte} of RAM.
On the SKL platform, this run achieves a performance of $\num{1.5}$ TFlop/s,
which, due to a rather moderate arithmetic intensity of the algorithm
(\num{0.3}), is bounded by the memory bandwidth. When increasing the number of
cores for a fixed problem size, the computations of the FFT and linear terms
show very good strong scalability, whereas the global transposes ultimately
limit the total parallel efficiency at large core counts
(Figure~\ref{fig:runtime}). A comprehensive study of the parallel scalability
and efficiency of \nsc has been presented in~\cite{Shi2015}, and its potential
to scale up to extremely high core counts was shown in~\cite{Rampp2014}. The
upgraded version presented here exhibits the same scalability and maintains
consistent performance over a range of different HPC systems without the need
for any platform-specific adaptations of the source code.
\par
\begin{figure}[htb]
\centering
\includegraphics[scale=1.00]{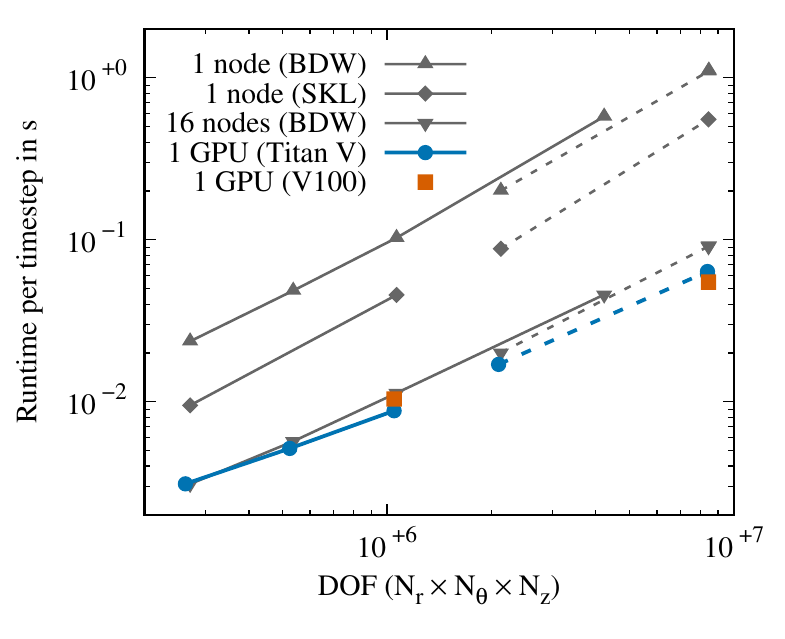}
\caption{Performance of the GPU-accelerated version of \nsc compared to the
\mpi-\omp version. Runtime per timestep for different numbers of degrees of
freedom (DOF). The GPU code ran on a single NVIDIA Titan V and a single Tesla
V100 graphics card. It was built using NVIDIA's \cuda toolkit version 10.1. The
hybrid code was built using Intel's \code{PSXE2018} and ran on one and \num{16}
nodes of different platforms. BDW: Intel Xeon E5-2620v4 (Broadwell), \num{16}
cores/node, Mellanox InfiniBand FDR network (\SI{56}{\giga\bit\per\second}).
SKL: Intel Xeon 6148 (Skylake), \num{40} cores/node. Solid (dashed) lines
represent runs with $N_r=\num{64}$ ($N_r=\num{128}$) radial points.}
\label{fig:scalingCompare}
\end{figure}

The performance of the GPU-accelerated version was tested on two NVIDIA graphics
cards based on the Volta architecture: Titan V and Tesla V100
(Figure~\ref{fig:scalingCompare}). The runtime per timestep has been found to be
similar on both cards over a wide range of problem sizes.
The speed-up of the GPU version compared to the HPS version running on one node
was shown to vary between a factor of three and \num{17},  depending on the
problem size and the particular choice of platform used as reference. Comparing
a single GPU run against an \mpi-\omp run on a single CPU node is a reasonable
choice, since for server-class hardware both set-ups are roughly comparable in
terms of price and electrical power consumption. However, \num{16} nodes
(\num{256} cores) were necessary to outperform the GPU version for small problem
sizes. Currently, the maximum problem size applicable to the GPU version is
limited by the amount of RAM available on the graphics card.

\subsection{Data analysis and visualization}
\label{sec:dataAnalysisAndVisualization}
% 150 words

In \nsc, the spectral coefficients and the primitive variables are dumped to
individual files for each timestep at user-specified output intervals. It
implements an easy-to-use checkpoint-restart mechanism based on the coefficients
for handling long-running DNS. The primitive variables -- velocity ($u_r$,
$u_{\theta}$, $u_{z}$), pressure ($p$) and optionally temperature ($T$) -- are
written in \code{HDF5} format, along with metadata in small \xmf files in order 
to facilitate analysis with common visualization software like \paraview and
\visit. Both tools allow loading sequences of \xmf files produced by \nsc.
Sample scripts based on the \python interface of \visit, as well as a
custom-made \paraview filter for handling the cylindrical coordinate system are
distributed with the code. A detailed visualization tutorial is included in the
user guide. This enables the user to easily perform comprehensive visual
and quantitative analysis of the flow field.

\subsection{Quality assurance}
\label{sec:qualityAssurance}
% 175 words

Verification and validation (V\&V) of \nsc is documented in \cite{Shi2015}. For
maintaining the correctness of the source code, we make extensive use of the
continuous integration (CI) functionality of \code{gitlab}. Upon every push to
the repository, a number of regression tests are automatically triggered,
including builds of the code in various configurations and a static code
analysis using the \code{Forcheck} tool. In addition, a number of short test
runs are automatically launched using runtime-checks and the tightest debug
settings of the compiler to identify undefined variables, out-of-bounds errors
and alike. The numerical results are then rigorously verified against previously
recorded reference runs. A final validation run compares the wave speed of a
simulated wavy vortex flow with an experimentally determined
value~\cite{King1984}, which is considered successful if the wave speeds match
up to \num{e-4}. The entire CI configuration, the V\&V results for every push, and an
auto-generated source code documentation (\ford) -- including dependencies
and call graphs -- are publicly accessible through the web interface of our
\href{https://gitlab.mpcdf.mpg.de/mjr/nscouette}{development site.}

\section{Illustrative Example}
\label{sec:illustrativeExample}
% 486 words

The fluid flow between a hot rotating cylinder and a cooled stationary enclosure
is a simple model to investigate heat transfer in many engineering
applications~\cite{Ali1990} such as cooling of rotating
machinery~\cite{Howey2012}. At low angular speeds (\Rei) and small temperature
differences (\Rayleigh), the heat transfer is purely conductive. In this simple
case, termed basic state, the governing equations admit analytic solutions for
$u_{\theta}$ and $T$, which only depend on $r$. The heat transfer can be
enhanced by either increasing \Rei (forced convection) or by increasing
\Rayleigh (natural convection). In both cases, the basic state exhibits a
sequence of distinct instabilities ultimately leading to turbulent heat
transfer~\cite{Lopez2015}. A measure of the efficiency is given by the Nusselt
number (\Nui), which is the ratio of total heat transfer at the inner cylinder,
normalized by that of the basic state at the same temperature difference.
\par
\begin{figure}[htb]
\centering
\includegraphics[width=1.00\linewidth]{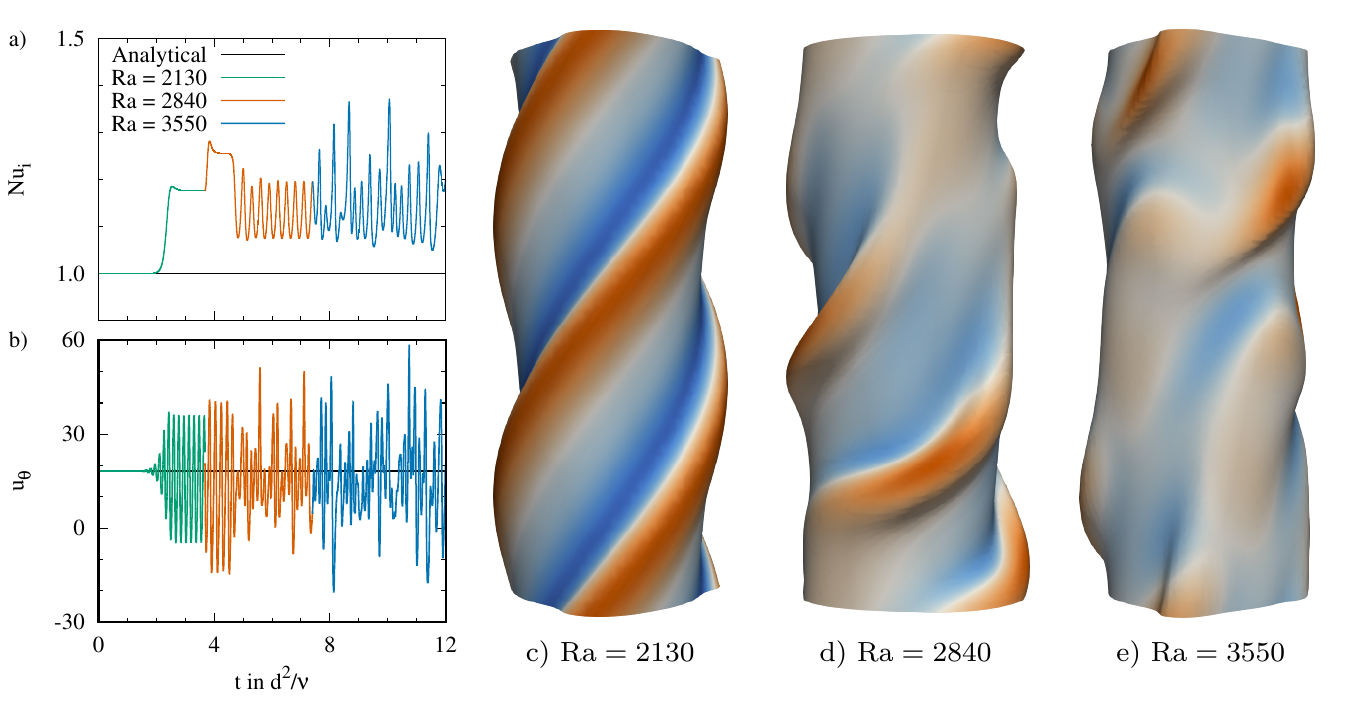}
\caption{Temporal evolution of the Nusselt number (\Nui) at the inner cylinder
wall and the streamwise velocity component ($u_{\theta}$) at a mid-gap position
as \Rayleigh increases for a fixed inner cylinder rotation ($\Rei=\num{50}$).
The final flow state for each \Rayleigh is visualized with instantaneous
temperature iso-surfaces ($T=0$), which are color-coded by inwards/outwards
(blue/red) facing values of the wall-normal velocity component ($u_r$).}
\label{fig:compareGr}
\end{figure}

Figure~\ref{fig:compareGr} summarizes the results of three DNS with increasing
temperature difference at a fixed rotation rate, using \nsc.
The first DNS was initialized by applying small single harmonic disturbances to
the basic state at $\Rei=\num{50}$ and $\Rayleigh=\num{2130}$.
Figure~\ref{fig:compareGr}a shows that initially $\Nui\approx\num{1}$,
corresponding to purely conductive heat transfer. However, after roughly two
viscous time units, a sharp increase of \Nui is observed, indicating that the
basic state has become unstable. This is confirmed by the time-series of
$u_{\theta}$ at a fixed probe location in the computational domain 
(Figure~\ref{fig:compareGr}b). The final state of this run is visualized in 
figure~\ref{fig:compareGr}c. It shows a three-dimensional rendering of a $T=0$
iso-surface generated with \paraview. The spiral flow pattern rotates at a
constant speed without changing its shape; like a barber pole. It passes
repeatedly through the probe location, what explains why the $u_{\theta}$
signal reaches a periodic state ($\num{2.5}\sfrac{d^2}{\nu}>t>\num{3.5}\sfrac{d^2}{\nu}$).
Its constant shape explains why the integral heat flux (\Nui), on the other hand,
remains constant at the same time. The second and third DNS were initialized with
the final state of the former runs and by increasing the Rayleigh number to
$\Rayleigh=\num{2840}$ and \num{3550}, respectively. The time series and final
states in figure~\ref{fig:compareGr} reveal, that the TC system undergoes a
sequence of transitions to different flow states with increasing spatio-temporal
complexity as \Rayleigh increases. These, and other illustrative examples, are
documented in the tutorial section of the user guide.

\section{Impact}
\label{sect:impact}
% 332 words

Our software completes the list of publicly available Navier--Stokes solvers for
the three most common prototypes of wall-bounded shear flow: plane Couette flow
(\cf~\cite{Gibson2014}), pipe flow (\opf~\cite{Willis2017}) and Taylor--Couette
flow (\nsc). It can be quickly installed and productively used by researchers
interested in pattern formation and chaos, for which TC flow has long been a
paradigm~\cite{Fardin2014}. The example of section~\ref{sec:illustrativeExample}
can be run in a laptop and is meant to illustrate how easy results can be
obtained and analyzed.
\par
We however stress that the main aim of providing \nsc is to make the full
potential of a massively-parallel and highly-scalable DNS code readily usable
and, therefore, valuable for the broad scientific community. It enables users
with little experience in HPC and DNS to easily perform highly-resolved
simulations of turbulence. Among other software design choices, this was
achieved by providing a user friendly build process that supports many
pre-configured HPC architectures, easy runtime handling of all control
parameters, tailored visualization tools, a comprehensive user guide and
thorough internal quality assurance. Thus, \nsc is a powerful tool to study
%angular momentum transport, mixing and heat transfer in
highly-turbulent shear flows. For example, it has already contributed to
a better understanding of astrophysical~\cite{Shi2017} and
geophysical~\cite{Leclercq2016a, Leclercq2016b} flows.
\par
Because of its modular structure, which closely follows the numerical
formulation, and its moderate code complexity (in particular the HPS),
new functionalities can be easily added to \nsc, with given algorithmic
domain knowledge and basic programming skills. As an example for this, we also
provide \nsp; a modified version to simulate flows in straight pipes. It
follows the numerical formulation of \opf~\cite{Willis2017} and uses the
parallel infrastructure of \nsc, as detailed in the user guide.
Extensions including the modelling of polymer additives~\cite{Lopez2019}
and two-phase flows~\cite{Song2019} have already been developed and tested
and will be released in the future.

\section*{Acknowledgements}
% 121 words

This work was supported by the Max Planck Society and partially funded by the
German Research Foundation (DFG) through the priority programme
\href{https://www.tu-ilmenau.de/turbspp/}{\textit{Turbulent Superstructures}
(SPP1881)}. AVM was supported by the European Research Council (ERC) through
the \textit{COTURB} project (ERC-2014.AdG-669505). Computational resources
provided by the following institutions are gratefully acknowledged: The Argonne
Leadership Computing Facility, which is a DOE Office of Science User Facility
(DE-AC02-06CH11357). The \href{http://gw4.ac.uk/isambard}{Isambard UK
National Tier-2 HPC Service} operated by GW4 and the UK Met Office, and funded
by EPSRC (EP/P020224/1). Further computations were performed on the HPC systems
Hydra, Draco and Cobra at the MPCDF in Garching.

\section*{References}

\bibliographystyle{elsarticle-num}
\bibliography{softx2020.bib}

\section*{Required Metadata}
\label{sect:sw_metadata}

\section*{Current code version}
\label{sect:code_version}

\begin{table}[!ht]
\begin{tabular}{|l|p{6.5cm}|p{6.5cm}|}
\hline
\textbf{Nr.} & \textbf{Code metadata description} &
\textbf{Please fill in this column} \\
\hline
C1 & Current code version & \code{1.0} \\
\hline
C2 & {\color{red}Permanent link} to code/repository used for this code version &
\url{https://github.com/dfeldmann/nsCouette} \\
\hline
C3 & Legal Code License & GPLv3\\
\hline
C4 & Code versioning system used & \code{git} \\
\hline
C5 & Software code languages, tools, and services used &
\fortran,
\c (for some housekeeping tasks),
\mpi,
\omp \\
\hline
C6 & Compilation requirements, operating environments \& dependencies &
Developed and tested under Linux and IBM AIX. Compiler:
A \fortran 2003 compiler which is \code{OpenMP-3} compliant,
a basic \c compiler,
an \mpi library with support for \code{MPI\_THREAD\_SERIALIZED},
a serial \code{BLAS}/\code{LAPACK} library,
a serial but fully thread-safe \code{FFTW3} installation or equivalent,
for output and visualization (optional): an \mpi-parallel \code{HDF5} installation.\\
\hline
C7 & Link to developer documentation/manual &
\url{https://gitlab.mpcdf.mpg.de/mjr/nscouette} \\
\hline
C8 & Support email for questions &
\url{nsCouette@zarm.uni-bremen.de} \\
\hline
\end{tabular}
\caption{Code metadata (mandatory)}
\label{}
\end{table}

\end{document}